\documentstyle[aps,eqsecnum,amssymb,amsbsy,preprint,tighten,epsfig]{revtex}
\title{Bounding the graviton mass with binary pulsar 
observations}
\author{
Patrick J. Sutton\thanks{Also Center for Gravitational Physics and 
Geometry and Department of Physics; e-mail psutton@gravity.phys.psu.edu} 
and 
Lee Samuel Finn\thanks{Also Center for 
Gravitational Physics and Geometry, Department 
of Physics, and Department of Astronomy and Astrophysics; e-mail 
lsf@gravity.phys.psu.edu} 
}       
\address{
Center for Gravitational Wave Physics, 
The Pennsylvania State University,
State College, PA, USA 16802-6300.
}

\begin{document}


\maketitle
\begin{abstract}
By comparing the observed orbital decay of the binary pulsars 
PSR~B1913+16 and PSR~B1534+12 to that predicted by general 
relativity due to gravitational-wave emission, we are able to bound 
the mass of the graviton to be less than $7.6\times10^{-20}\,\text{eV}/c^2$ 
at $90\%$ confidence.  This is the first such bound to be derived 
from dynamic gravitational fields.  It is approximately two 
orders of magnitude weaker than the static-field bound from solar system 
observations, and will improve with further observations.
\end{abstract}


\section{Introduction}
\label{Introduction}

General relativity assumes that the graviton has zero rest mass. 
For static gravitational fields a nonzero graviton mass $m$ 
would cause the potential to tend to the Yukawa form $r^{-1} e^{-mr}$, 
effectively cutting off gravitational interactions at distances 
larger than the Compton wavelength $m^{-1}$ of the graviton. 
Current experimental limits on the graviton mass are based on 
the apparent absence of such a cutoff in the solar system 
\cite{talmadge88a} and in galaxy and cluster dynamics 
\cite{goldhaber74a,hare73a}. 

The study of dynamic gravitational fields allows new and 
independent limits to be placed on the graviton mass.  For example, 
Will \cite{will98a} and Larson and Hiscock \cite{larson00a} 
have shown how future observations 
of gravitational waves may be used to bound the 
graviton mass via comparison to the dispersion formula 
\begin{equation}\label{speed}
v(\omega) = \sqrt{1-\frac{m^2}{\omega^2}} \, .
\end{equation}
Here we propose a new technique for limiting $m$ 
using available data on the orbital decay 
of binary pulsars.  Consider the Hulse-Taylor 
binary pulsar, PSR~B1913+16, for which the observed orbital decay 
attributed to gravitational-wave emission agrees with the predictions 
of general relativity to approximately $0.3\%$.  A nonzero graviton 
mass would alter the energy emission rate\footnote{
Corrections to other characteristics of the system are 
negligible by comparison on dimensional grounds: 
$(mr)^2=(m/\omega)^2 (v/c)^2$, $(mM)^2=(m/\omega)^2(v/c)^6$, 
where $v/c = {\cal O}(10^{-3})$ for these binary systems.
} 
and destroy this agreement; 
for gravitational waves at twice the orbital frequency of PSR~B1913+16,  
requiring 
\begin{equation}
\left(\frac{m}{\omega}\right)^2 < 0.003 
\end{equation}
implies $m < {\cal O}(10^{-20})\,\text{eV}/c^2$.  This is comparable to the limit  
$m < 4.4\times10^{-22}\text{eV}/c^2$ obtained from solar system observations 
\cite{talmadge88a}.  

In this paper we develop this bound in more detail.  
In Section~\ref{framework} we examine linearized general relativity 
with a massive graviton.  In Section~\ref{solutions} we calculate 
the corrections to the energy emission rate of a compact, slowly moving 
source due to a nonzero graviton mass.  Comparison to the observed orbital 
decay of  PSR~B1913+16 and PSR~B1534+12 will provide us with an upper limit  
on the graviton mass in Section~\ref{binarypulsars}.  We conclude in 
Section~\ref{conclusions} with a few brief comments.


\section{Linearized General Relativity with a Massive Graviton}
\label{framework}

Gravitational waves on a flat background spacetime can be described as 
a perturbation of the Minkowski metric:   
\begin{equation}\label{h}
g_{\mu\nu} = \eta_{\mu\nu} + h_{\mu\nu} \, , \qquad |h_{\mu\nu}| \ll 1 \, . 
\end{equation}
For convenience we work with the trace-reversed metric, 
\begin{equation}
\bar{h}_{\mu\nu} 
  \equiv  h_{\mu\nu}-{1\over 2}\eta_{\mu\nu} h^\lambda{}_\lambda \, ,
\end{equation}
and raise and lower indices using $\eta^{\mu\nu}$, $\eta_{\mu\nu}$. 
Substituting (\ref{h}) into the Einstein field equations, 
expanding in powers of $\bar{h}_{\mu\nu}$, and keeping only 
terms up to first order yields   
\begin{equation}\label{masslessfieldeqn}
\Box \bar{h}_{\mu\nu} =  -16\pi \,T_{\mu\nu}  \, ,
\end{equation}
where the stress tensor $T_{\mu\nu}$ of the source matter is conserved, 
$T_{\mu\nu}^{\hphantom{\mu\nu},\nu} = 0$, and we work in the 
Lorentz gauge, defined by $\bar{h}^{\mu\nu}{}_{,\nu} = 0$.

A simple generalization of (\ref{masslessfieldeqn}) to include a 
phenomenological mass for the graviton is \cite{visser98a} 
\begin{equation}\label{fieldeqn}
(\Box-m^2) \bar{h}_{\mu\nu} =  -16\pi \,T_{\mu\nu}  \, .
\end{equation}
The massive field $\bar{h}_{\mu\nu}$ has six degrees of freedom: 
five spin-2, and one scalar \cite{visser98a,boulware72a}.  
It is not gauge invariant, 
but the Lorentz condition $\bar{h}^{\mu\nu}{}_{,\nu} = 0$ still 
holds as a constraint. 

An effective stress tensor for the gravitational waves can be 
obtained using Noether's theorem \cite{wald84a}, and is 
identical in form to the usual $m=0$ result \cite{misner73a}:
\begin{equation}\label{canonical1}
T^{\text{GW}}_{\mu\nu}
  =  \frac{1}{32\pi}\langle 
         \bar{h}_{\alpha\beta,\mu} \bar{h}^{\alpha\beta}{}_{,\nu}
         -\frac{1}{2}\bar{h}^\alpha{}_{\alpha,\mu}\bar{h}^\beta{}_{\beta,\nu}
     \rangle \, . 
\end{equation}
Here the brackets denote an averaging over at least one period of the 
gravitational wave.


\section{Energy Emission Rate}
\label{solutions}

We now calculate the energy emission rate of compact, slowly moving 
sources of gravitational waves described by 
(\ref{fieldeqn})--(\ref{canonical1}).  Our analysis follows 
that used for standard linearized general relativity \cite{misner73a}, 
except that we work in the frequency domain for convenience \cite{krauss94a}.  

For a periodic source with period $P$, we decompose the 
metric perturbation as 
\begin{equation}\label{sum} 
\bar{h}_{\mu\nu}(t,\vec{x}) 
  =  \sum_{n=-\infty}^\infty \widetilde{\bar{h}}_{\mu\nu}(\omega_n,\vec{x}) 
     \, e^{-i\omega_n t} \, , 
\end{equation}
where 
\begin{equation}
\omega_n = n \, \frac{2\pi}{P} \, 
\end{equation}
with $n$ an integer.  Then (\ref{fieldeqn}) becomes
\begin{equation}\label{FTfieldeqn}
\left(\nabla^2+[\omega^2-m^2]\right)
\widetilde{\bar{h}}_{\mu\nu}(\omega | \vec{x}) 
  =  -16 \pi \,\widetilde{T}_{\mu\nu}(\omega | \vec{x}) \, ,
\end{equation}
where $\nabla^2$ is the 3-space Laplacian.  The retarded Green function for 
this equation is
\begin{equation}\label{FTG}
\widetilde{G}_R(\omega|\vec{x};\vec{x}') 
  =  \frac{e^{ik|\vec{x}-\vec{x}'|}}{4\pi |\vec{x}-\vec{x}'|}  \, ,
\end{equation}
where $k\equiv \mbox{sign}(\omega)\sqrt{\omega^2-m^2}$ for $|\omega| > m$. 
The retarded solution of (\ref{FTfieldeqn}) for fixed $\omega$ is then 
\begin{equation}\label{FTgensoln}  
\widetilde{\bar{h}}_{\mu\nu}(\omega | \vec{x}) 
  =  16\pi \int\!d^3\!x' \, \widetilde{G}_R(\omega|\vec{x};\vec{x}') \,
     \widetilde{T}_{\mu\nu}(\omega | \vec{x}') \, .
\end{equation}
Using the slow-motion approximation ($\omega a \ll 1$, with $a$ the 
characteristic size of the source), taking the observation point far 
from the source region ($r\equiv|\vec{x}| \gg a$), and employing the 
conservation of the stress tensor, one obtains  
\begin{eqnarray}
\widetilde{\bar{h}}_{00}(\omega | \vec{x})
  & = &  \frac{4e^{ikr}}{r} \left[ \, 
             \widetilde{M} 
             +\frac{x^j}{r} (-ik) \widetilde{D}_{j} 
	     +\frac{x^jx^k}{2r^2}(-ik)^2 \widetilde{I}_{jk}
	 \, \right]  , \nonumber \\
\widetilde{\bar{h}}_{0j}(\omega | \vec{x})
  & = &  \frac{4e^{ikr}}{r} \left[ \, 
             -(-i\omega) \widetilde{D}_{j} 
	     -\frac{x^k}{2r}(-ik)(-iw) \widetilde{I}_{jk}
	 \, \right]  , \nonumber \\
\widetilde{\bar{h}}_{jk}(\omega | \vec{x}) 
  & = &  \frac{4e^{ikr}}{r} \left[ \, 
             \frac{1}{2} (-i\omega)^2 \widetilde{I}_{jk}  
	 \, \right]  , \label{soln} 
\end{eqnarray}
where $\widetilde{M}$, $\widetilde{D}_j$, $\widetilde{I}_{jk}$, are 
respectively the Fourier coefficients
of the mass, dipole moment, and quadrupole moment of the source.  
Only the quadrupole terms contain nonzero-frequency 
components and contribute to the radiation. 

Substituting (\ref{soln}) into the effective stress tensor 
(\ref{canonical1}) for the gravitational waves and integrating 
the outward flux over a sphere centered on the source gives the 
rate of energy emission.  One finds 
\begin{mathletters}\label{eqn:Eloss} 
\begin{equation}
    L \equiv -\frac{d E}{d t} 
     =  L_{\text{GR}} + \sum_{n=1}^\infty 
    \frac{m^2 \omega_n^4}{3} \left[
             \widetilde{I}_{jk}(\omega_n)\widetilde{I}_{jk}^*(\omega_n) 
	     -\left|\text{tr}\,\widetilde{I}(\omega_n)\right|^{2}
         \right]
         + {\cal O}\left(m^4\right) \, ,\label{eqn:Lm}
\end{equation}
where 
\begin{equation}
    L_{\text{GR}}  \equiv 
    \sum_{n=1}^\infty 
    \omega_n^6 \left[
             \frac{2}{5}\widetilde{I}_{jk}(\omega_n)
	         \widetilde{I}_{jk}^*(\omega_n) 
	     -\frac{2}{15}\left|\text{tr}\,\widetilde{I}(\omega_n)\right|^{2}
         \right] 
\end{equation}
\end{mathletters}
is the usual general-relativistic expression for the radiated power,  
$\text{tr}\,\widetilde{I}$ is the trace of $\widetilde{I}_{jk}$, and
we sum over repeated indices.  Equation (\ref{eqn:Lm}) gives the 
corrections to the radiated power due to a small nonzero graviton mass; 
comparison to the observed orbital decay in binary pulsars 
PSR~B1913+16 and PSR~B1534+12 will provide us with a bound on $m$.


\section{Binary Pulsars}
\label{binarypulsars}

The formula (\ref{eqn:Eloss}) for the energy-loss rate of a
gravitational-wave source when the graviton is massive is easily
applied to the orbital decay of binary systems to put a limit on 
the graviton mass.  Consider PSR~B1913+16, for which 
the orbital decay rate is slightly in excess of the predictions of general
relativity \cite{taylor94a}.  
Denote by $P_{\text{b}}$ the measured orbital period of the binary system, 
$\dot{P}_{\text{b}}$ the measured orbital period derivative ascribed to 
gravitational radiation, and $\dot{P}_{\text{GR}}$ the instantaneous 
period derivative expected from general relativity.
For a slowly decaying Keplerian binary, the instantaneous period derivative 
is proportional to the energy-loss rate; hence, 
\begin{equation}\label{fracdisc}
\Delta 
\equiv \frac{\dot{P}_{\text{b}}-\dot{P}_{\text{GR}}}{\dot{P}_{\text{GR}}} 
  =  \frac{L - L_{\text{GR}}}{L_{\text{GR}}} \, ,
\end{equation}
where $L$ is the gravitational-wave luminosity inferred from 
$\dot{P}_{\text{b}}$, and $L_{\text{GR}}$ is the energy-loss rate 
expected from general relativity.  This fractional discrepancy $\Delta$ 
has been measured for PSR~B1913+16 and PSR~B1534+12 
(see \cite{taylor94a,stairs99a} and Table~\ref{tbl:pulsar}).

\begin{figure}
    \epsfxsize=\columnwidth
    \epsfxsize=5in
    \epsffile{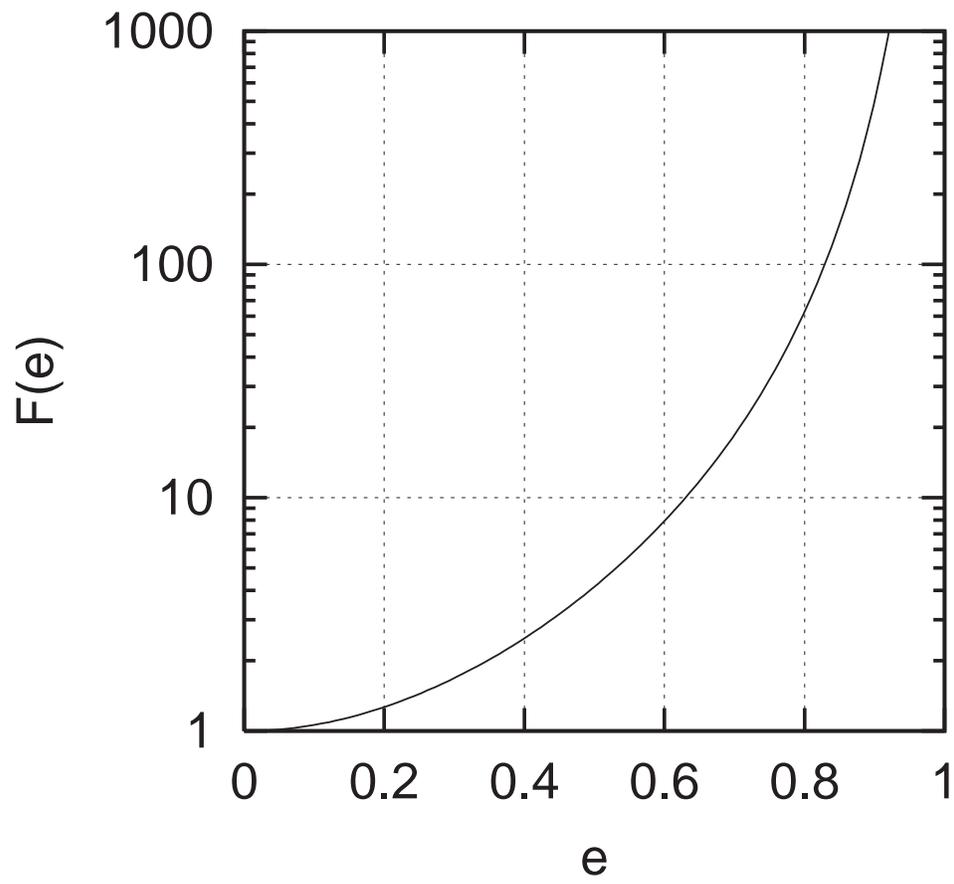}
\caption{
Eccentricity factor $F(e)$ (cf.\ eqn.\ \ref{eq:F(e)summed}) versus $e$.
}
\label{fig:F}
\end{figure}

Now suppose that $\Delta$ is due at least in part to a
nonvanishing graviton mass.  Combining (\ref{eqn:Eloss}) 
and (\ref{fracdisc}) implies 
\begin{equation}\label{bound}
m^{2} 
  \le  \frac{24}{5}\,F(e)\,
       \left(\frac{2\pi\hbar}{c^{2}P_{\text{b}}}\right)^{2} \, 
       \frac{\dot{P}_{\text{b}}-\dot{P}_{\text{GR}}}{\dot{P}_{\text{GR}}} \, ,
       \label{eqn:dl/l}
\end{equation}
where $F(e)$ is a function of the eccentricity,
\begin{equation}
F(e)
  \equiv  \frac{1}{12}\frac{
     \displaystyle{\sum_{n=1}^\infty} \,n^6 \left[
         3\widetilde{I}_{jk}(\omega_n)\widetilde{I}_{jk}^*(\omega_n)
         -\left|\text{tr}\,\widetilde{I}(\omega_n)\right|^{2}
     \right]
     }{
     \displaystyle{\sum_{n=1}^\infty} \,n^4 \left[
         \widetilde{I}_{jk}(\omega_n)\widetilde{I}_{jk}^*(\omega_n)
         -\left|\text{tr}\,\widetilde{I}(\omega_n)\right|^{2}
     \right]
     }  
  =  \frac{1+\frac{73}{24}e^2+\frac{37}{96}e^4}{(1-e^2)^3}
     \, , \label{eq:F(e)summed}
\end{equation}
as can be shown using the techniques of \cite{peters63a}; 
see Figure~\ref{fig:F}.  

Equations (\ref{bound}), (\ref{eq:F(e)summed}) show that a nonzero 
graviton mass increases the energy emission and decay rate 
of Keplerian binaries, as one would expect from adding extra degrees of 
freedom to the gravitational field.  The strongest bounds arise from
binaries with small eccentricity and large period, as these systems 
produce the bulk of their radiation at low frequencies \cite{peters63a},  
which are the most sensitive to a graviton mass, as in equation
(\ref{speed}).  

In using (\ref{eqn:dl/l}) to place an upper limit on the graviton 
mass, we should take into account the experimental uncertainties 
in the fractional discrepancy $\Delta$, which are typically of the 
same order as $\Delta$.  We assume the
measured discrepancy $\Delta$ to be normally distributed about its unknown 
actual value (given by the equality in (\ref{bound}) with unknown $m^2$), 
and with standard deviation as given in Table \ref{tbl:pulsar}.  In our
model we relate the discrepancy to the squared graviton mass, which
must be non-negative.  Referring to \cite[Table X]{feldman98a}, which
lists the 90\% unified upper limit/confidence intervals for the
non-negative mean of a univariate normal distribution 
based on a measured sample from the distribution, we calculate the
90\% upper limit on the graviton mass, which is given
in the final row of Table \ref{tbl:pulsar}.
These two observations of $m^2$ may also be combined into a single upper 
bound on the graviton mass by averaging the individual $m^2$ bounds 
with weight according to their variances.  Again referring to 
Table \ref{tbl:pulsar} and \cite[Table X]{feldman98a}, 
the corresponding limit on the graviton mass from the combined 
observations of PSR~B1913+16 and PSR~B1534+12 is found to be 
\begin{equation}
    m_{90\%} < 7.6\times10^{-20}\,\text{eV}/c^2.
\end{equation}

\section{Discussion}
\label{conclusions}

Table \ref{tbl:pulsar} gives the relevant parameters and the
corresponding graviton mass bounds for the two binary pulsars whose
gravitational-wave induced orbital decay has been measured, PSR~B1913+16 
and PSR~B1534+12 \cite{taylor94a,stairs99a}.  These bounds are about 
two orders of magnitude weaker than the Yukawa
limit obtained from solar-system observations, $mc^2 < 4.4 \times
10^{-22}\,\text{eV}$ \cite{talmadge88a}, and 
several orders of magnitude weaker than that provided by observations
of galactic clusters, $mc^2 < 2 \times 10^{-29}\,\text{eV}$
\cite{goldhaber74a,hare73a}, though these galactic cluster
bounds may be less robust, owing to their reliance on assumptions about
the dark matter content of the clusters.  In contrast,
the bound obtained here is very straightforward.  Our chief assumption  
is the form of the effective mass term for the graviton,
which, while not unique, is natural.  Furthermore, any other
mass term would be expected from dimensional arguments to yield
similar results.

Our other major assumption is that only unbiased measurement errors 
enter into the determination of the intrinsic binary period decay 
rate $\dot{P}_{b}$. 
The determination of $\dot{P}_{b}$ requires an accurate distance 
measurement to the binary system, however, which can be difficult to make.  
The large uncertainty
in the discrepancy $\Delta$ associated with PSR~B1534+12 may well be
due to an underestimate of the distance to this binary system
\cite{stairs99a}, in which case the bound on $m^{2}$ would be even
tighter.

The bound described here arises from the properties of dynamical
relativity, making it conceptually independent of either the solar
system or galactic cluster bounds on the graviton mass, which
are based on the Yukawa form of the static field in a massive theory. 
Furthermore, the bounds from any given pulsar system will improve 
as observations increase the accuracy of the measured fractional 
discrepancy in the period derivative.  

 
\section*{Acknowledgments}

The authors are grateful to Valeri Frolov, Matt Visser, Cliff Will, 
Alex Wolszczan, and Andrei Zelnikov for helpful discussions.  PJS would
like to thank the LIGO Scientific Collaboration and the 
Natural Sciences and Engineering Research Council of
Canada for their financial support.  This work has been funded by NSF
grant PHY~00-99559 and its predecessor.  The Center for Gravitational
Wave Physics is supported by the NSF under co-operative agreement
PHY~01-14375.



\begin{table}[b]
\caption{Orbital parameters and corresponding graviton mass bound 
from the two binary pulsar systems whose gravitational wave 
induced orbital decay has been measured. Pulsar parameters are 
taken from \protect\cite{taylor94a,stairs99a}. One-sigma uncertainties are 
quoted for $\Delta$.}\label{tbl:pulsar}  
\begin{tabular}{l|rr}
               & PSR~B1913+16           & PSR~B1534+12 \\
  \hline
  Period       & 27907~s                & 36352~s \\
  Eccentricity & 0.61713                & 0.27368 \\
  $\Delta$     & 0.32\% $\pm$ 0.35\%    & $-12.0$\% $\pm$ 7.8\% \\
  Graviton mass 90\% upper bound& $9.5\times10^{-20}\,\text{eV}/c^2$ & 
  $6.4\times10^{-20}\,\text{eV}/c^2$ \\
\end{tabular}
\end{table}

\end{document}